\documentclass[12pt]{article}

\usepackage[totalwidth=460truept,totalheight=610truept]{geometry}
\usepackage{amsfonts,latexsym,amssymb,amsmath,graphicx,accents,slashed,subfigure}
\usepackage{latexsym}
\usepackage[hidelinks]{hyperref}

\global\arraycolsep=1truept

\begin{document}

\null

\bigskip \phantom{C}

\begin{center}
{\huge \textbf{Fakeons, Unitarity,}}

\vskip.5truecm

{\huge \textbf{Massive Gravitons}}

\vskip.7truecm

{\huge \textbf{And The Cosmological Constant}}

\vskip 1truecm

\textsl{Damiano Anselmi}

\vskip .2truecm

\textit{Dipartimento di Fisica ``Enrico Fermi'', Universit\`{a} di Pisa, }

\textit{Largo B. Pontecorvo 3, 56127 Pisa, Italy}

\textit{and INFN, Sezione di Pisa,}

\textit{Largo B. Pontecorvo 3, 56127 Pisa, Italy}

\vskip.2truecm

damiano.anselmi@unipi.it

\vskip 1.5truecm

\textbf{Abstract}
\end{center}

We give a simple proof of perturbative unitarity in gauge theories and
quantum gravity using a special gauge that allows us to separate the
physical poles of the free propagators, which are quantized by means of the
Feynman prescription, from the poles that belong to the gauge-trivial
sector, which are quantized by means of the fakeon prescription. The proof
applies to renormalizable theories, including the ultraviolet complete
theory of quantum gravity with fakeons formulated recently, as well as
low-energy (nonrenormalizable) theories. We clarify a number of subtleties
related to the study of scattering processes in the presence of a
cosmological constant $\Lambda $. The scattering amplitudes, defined by
expanding the metric around flat space, obey the optical theorem up to
corrections due to $\Lambda $, which are negligible for all practical
purposes. Problems of interpretation would arise if such corrections became
important. In passing, we obtain local, unitary (and \textquotedblleft
almost\textquotedblright\ renormalizable) theories of massive gravitons and
gauge fields, which violate gauge invariance and general\ covariance
explicitly.

\vfill\eject

\section{Introduction}

\label{s0}

\setcounter{equation}{0}

Unitarity, i.e. the statement that the $S$ matrix satisfies $S^{\dag }S=1$,
is a key principle of perturbative quantum field theory, together with
locality and renormalizability. It can be proved diagrammatically by means
of the so-called cutting equations \cite{cutkosky,diagrammar}, which are
sums of diagrams made of two parts, one associated with $S$ and the other
associated with $S^{\dag }$. The proof is relatively straightforward in
theories of scalar fields and fermions. Gauge theories are more demanding,
since they require to show that the Faddeev-Popov ghosts and the
longitudinal and temporal components of the gauge fields mutually compensate
and can be projected away. A direct analysis of this compensation dates back
to the early '70s and is due to 't Hooft \cite{thooft}.

In this paper we study these issues by means of more modern techniques. The
first goal is to simplify and generalize the proof of perturbative unitarity
by using the concept of fake particle, or fakeon \cite{LWgrav,fakeons}. The
fakeon is a degree of freedom that can only be virtual and must be
consistently projected away from the physical spectrum to have unitarity.
The consistency of the fakeon projection does not follow from a gauge
principle, but from a new quantization prescription. Under certain
assumptions, fakeons can make sense of higher-derivative theories. They
provide a better understanding of the Lee-Wick models \cite{leewick} and
actually lead to the completion of their formulation \cite{Piva}, which had
ambiguities \cite{cutk} and issues related to Lorentz invariance \cite%
{nakanishi}. Fakeons can also be applied to non-higher-derivative theories
and allow us to formulate a consistent theory of quantum gravity \cite%
{LWgrav,UVQG,absograv}.

The fakeons are introduced by quantizing some poles of the free propagators
in momentum space by means of the fakeon prescription, which works as
follows:

($i$) at the tree level, the free fakeon propagator coincides with the Cauchy
principal value of the unprescribed propagator;

($ii$) inside the Feynman diagrams, the thresholds (and the cuts associated
with them) coincide with those determined by the Feynman prescription (or by
Wick rotating the Euclidean diagram), but they are bypassed in different
ways:

\qquad ($ii$-$a$) the thresholds associated with the processes that involve
at least one fakeon (which we call fake thresholds) are circumvented by
means of the average continuation \cite{Piva,fakeons}, which is the
arithmetic average of the two analytic continuations;

\qquad ($ii$-$b$) instead, the physical thresholds (those that do not
involve fakeons) are circumvented analytically, as usual.

The fakeon prescription is consistent with unitarity for every nonzero value
(positive, negative or complex) of the residue at the pole, as long as the
real part of the squared mass is nonnegative.

A special gauge \cite{unitarityc} allows us to separate the poles
corresponding to the physical helicities of the graviton and the gauge
fields from the poles that belong to the gauge-trivial sector. We quantize
the former by means of the Feynman prescription and the latter as fakeons,
i.e. by means of the fakeon prescription. A gauge-fixing parameter $\lambda $
is conveniently kept free. We use the $\lambda $ dependence inside the loop
diagrams to distinguish the physical thresholds, which are overcome
analytically, from the fake thresholds, which are overcome by means of the
average continuation. So doing, the proof of unitarity in gauge theories and
quantum gravity is relatively straightforward, comparable to the one of
scalar-fermion models. Our results apply to ordinary renormalizable
theories, low-energy (nonrenormalizable) effective theories, as well as the
ultraviolet complete theory of quantum gravity formulated in 2017 in ref. 
\cite{LWgrav}.

A nontrivial issue is due to the cosmological constant $\Lambda $, which
cannot be completely turned off in realistic models of quantum gravity. A
consistent formulation of the theory of scattering at $\Lambda \neq 0$ is
currently unavailable and might even not exist \cite{adsscat}. Yet, we know
that, when we make scattering experiments in our laboratories, we do not
care whether the universe has a cosmological constant or not. Since the
value of $\Lambda $ is very small in nature, we expect that its effects are
negligible for all practical purposes. While it is obvious that $\Lambda $
can be treated perturbatively at the classical level, it is not equally
obvious that we can do so in quantum field theory, where divergences can
simplify small quantities and return finite results.

We overcome these obstacles by showing that if we formulate the theory of
scattering in the presence of the cosmological constant by expanding around
flat space, perturbative unitarity holds up to corrections due to the
cosmological constant itself, which are indeed negligible for all practical
purposes. In the (unrealistic) situations where such corrections were not
negligible, our approach gives well-defined cutting equations, but does not
provide a physical interpretation for them in the realm of a theory of
scattering.

For various purposes, it is necessary to equip the gauge fields and the
graviton with small artificial masses, which we call gauge masses. Gauge
invariance, Lorentz invariance and general covariance are violated when the
gauge masses are nonzero and recovered when they are sent to zero. An
unexpected feature of our approach is that unitarity holds (up to the
corrections due to $\Lambda $) even when the gauge masses are nonvanishing.
Contrary to the common lore that gauge invariance and general covariance
cannot be explicitly broken without violating unitarity, the fakeons allow
us to achieve precisely that goal. Specifically, we can formulate theories
of massive gauge fields and gravitons that are local and unitary (in the
sense explained above). In the case of gauge fields, they are also
renormalizable. In the case of gravitons, they are \textquotedblleft
almost\textquotedblright\ renormalizable. We briefly compare such massive
theories with the known approaches to massive gravitons \cite%
{rubakov,dgp,drgt}.

The paper is organized as follows. In section \ref{specialgauge} we prove
unitarity in Yang-Mills theories. In section \ref{qg1} we extend the proof
to the low-energy (nonrenormalizable) theory of quantum gravity, at
vanishing cosmological constant. In section \ref{qed} we formulate the
theory of scattering in the presence of a cosmological constant. In section %
\ref{qg2} we extend the proof of unitarity to the ultraviolet complete
theory of quantum gravity of \cite{LWgrav} and its higher-dimensional
versions \cite{correspondence}. In section \ref{massive} we discuss the
properties of the theories of massive gravitons and gauge fields that emerge
from our approach. Section \ref{conclusions} contains the conclusions.
Whenever it is necessary to specify a regularization technique, we use the
dimensional one.

\section{Yang-Mills theories}

\label{specialgauge}

\setcounter{equation}{0}

In this section we prove unitarity in Abelian and non-Abelian gauge theories
in dimensions $d>2$, by quantizing the gauge-trivial sector with the fakeon
prescription. The generalization of the arguments to the coupling to matter
is straightforward (if the theory is manifestly anomaly free, which we
assume here), so we focus on pure gauge theories.

We start from four dimensions. Consider the gauge-fixed Lagrangian 
\begin{equation}
\mathcal{L}_{\mathrm{gf}}=-\frac{1}{4}F_{\mu \nu }^{a}F^{a\hspace{0.01in}\mu
\nu }-\frac{1}{2\lambda }\mathcal{G}^{a}(A)\mathcal{G}^{a}(A)-\bar{C}^{a}%
\mathcal{G}^{a}(DC)+\frac{m_{0}^{2}}{2}A_{0}^{a2}-\frac{m_{\gamma }^{2}}{2}%
\mathbf{A}^{a2}-\bar{m}^{2}\bar{C}^{a}C^{a},  \label{lmg}
\end{equation}%
where $A^{a\mu }=(A^{a0},\mathbf{A}^{a})$, $\lambda $ is a positive
gauge-fixing parameter, $C^{a}$ and $\bar{C}^{a}$ are the Faddeev-Popov
ghosts and antighosts, respectively, $D$ is the covariant derivative and $%
\mathcal{G}^{a}(A)$ denotes the gauge-fixing functions, which we assume to
be linear in $A$. Gauge masses $m_{0}$, $m_{\gamma }$, $\bar{m}$ are
included to regulate the on-shell infrared divergences of the cutting
equations. Lorentz invariance and gauge invariance are explicitly broken at
nonvanishing gauge masses. They are smoothly recovered in the limit of
vanishing gauge masses.

We work in the \textquotedblleft special gauge\textquotedblright\ of ref. 
\cite{unitarityc}, which amounts to take\footnote{%
Note that the mass terms of (\ref{lmg}) are slightly different from those of 
\cite{unitarityc}. Indeed, the approach of the present paper is more
versatile than the one of \cite{unitarityc} and allows us to make important
simplifications.} 
\begin{equation}
\mathcal{G}^{a}(A)=\lambda \partial _{0}A_{0}+\mathbf{\nabla }\cdot \mathbf{%
A.}  \label{gao}
\end{equation}%
The propagators derived from (\ref{lmg}) are then 
\begin{eqnarray}
\left\langle A^{0}(k)A^{0}(-k)\right\rangle _{0} &=&-\left. \frac{i}{\lambda
E^{2}-\mathbf{k}^{2}-m_{0}^{2}}\right\vert _{\text{f}},\qquad \qquad
\left\langle A^{i}(k)A^{0}(-k)\right\rangle _{0}=0,  \notag \\
\left\langle A^{i}(k)A^{j}(-k)\right\rangle _{0} &=&\frac{i\Pi ^{ij}}{E^{2}-%
\mathbf{k}^{2}-m_{\gamma }^{2}+i\epsilon }+\left. \frac{i\lambda }{\lambda
E^{2}-\mathbf{k}^{2}-\lambda m_{\gamma }^{2}}\right\vert _{\text{f}}\frac{%
k^{i}k^{j}}{\mathbf{k}^{2}},  \label{specialone} \\
\left\langle C(k)\bar{C}(-k)\right\rangle _{0} &=&\left. \frac{i}{\lambda
E^{2}-\mathbf{k}^{2}-\bar{m}^{2}}\right\vert _{\text{f}},  \notag
\end{eqnarray}%
where $k^{\mu }=(E,\mathbf{k})$ and 
\begin{equation}
\Pi ^{ij}=\delta ^{ij}-\frac{k^{i}k^{j}}{\mathbf{k}^{2}}  \label{proja}
\end{equation}%
is the transverse projector.

We have already inserted the quantization prescriptions we need.
Specifically, we quantize the physical poles (which are given by the
transverse components of $A^{i}$) by means of the Feynman $i\epsilon $
prescription and all the unphysical poles as fakeons, i.e. by means of the
fakeon prescription. The latter is denoted by the subscript
\textquotedblleft f\textquotedblright\ and, as recalled in the introduction,
amounts to circumvent the fake thresholds (those that involve at least one
fakeon) inside the loop diagrams by means of the average continuation. The
thresholds (and the cuts associated with them) coincide with those determined
by the Feynman prescription or by Wick rotating the Euclidean version
of the diagram. At the tree level, the free fakeon propagator coincides with the
principal value of the unprescribed propagator.

A virtue of the special gauge is that the propagators have only simple poles
for arbitrary $\lambda $. Instead, the usual Lorenz gauge-fixing function $%
\mathcal{G}(A)=\partial ^{\mu }A_{\mu }$ leads to double poles whenever $%
\lambda \neq 1$.

A caveat concerns the situations where two or more thresholds coincide,
which must be treated as limits of distinct thresholds \cite{fakeons}. For
example, the square of the principal value distribution is ill defined. But
if we split the singularities and make them coincide at the end, we obtain
(minus) the derivative of the principal value, which is well defined. See
details in ref. \cite{flrw}. This is the right method to evaluate the loop
integrals.

We proceed as follows. First,\ we deform the masses inside the loop diagrams
(independently for every propagator), to eliminate the coinciding
thresholds. In what follows, the parameter that measures this deformation
will be called $\eta $. Second, we complexify the external momenta $p$ and
compute the integrals in the Euclidean region, where the prescriptions are
immaterial (since no thresholds appear) and analyticity holds. Then we move
towards the subspace of real external momenta. When we do so, we find
physical and fake thresholds. The propagators (\ref{specialone}) ensure that
the physical thresholds are $\lambda $ independent, while the fake ones do
depend on $\lambda $. This allows us to keep them distinct and overcome them
in different ways. Specifically, the physical thresholds are circumvented by
means of the Feynman prescription, that is to say analytically. Instead, the
fake thresholds are circumvented by means of the average continuation, that
is to say by taking the arithmetic average of the two analytic
continuations. The average continuation is safe at $\eta \neq 0$, since
there are no coinciding thresholds by construction. At the end, we remove
the $\eta $ deformation by taking the limit $\eta \rightarrow 0$.

\subsubsection*{Renormalizability}

We recall that it is sufficient to prove the renormalizability of the theory
in the Euclidean framework \cite{LWgrav,fakeons}, because the average
continuation of convergent functions is obviously convergent. In other
words, the prescriptions do not affect the divergent parts of Feynman
diagrams.

The diagrams $G$ that contain coinciding thresholds are deformed as
explained above into diagrams $G_{\mathrm{split}}(\eta )$. The counterterms
are modified consistently. Once we subtract the subdivergences and the
overall divergence, we obtain a function 
\begin{equation}
G_{\mathrm{split}}(\eta )-\sum_{i}G_{\mathrm{split}}^{(i)\mathrm{sub}}(\eta
)-G_{\mathrm{split}}^{\mathrm{ovrll}}(\eta )  \label{subtra}
\end{equation}%
that is convergent in the Euclidean region. Then we move from the Euclidean
region to any other region, by taking the average continuation where
necessary. After that, we take the limit $\eta \rightarrow 0$. Clearly, the
result of these operations is convergent. It is easy to prove by using
standard tricks in the Euclidean region that every deformed counterterm is
polynomial in $\eta $ (which can be treated as a mass here) and tends to the
right counterterm for $\eta \rightarrow 0$.

A possible source of worry comes from the denominators $\mathbf{k}^{2}$. In
principle, they could lead to violations of the locality of counterterms
(see \cite{unitarityc} for details). In fact, they do not, because they
cancel out when the prescriptions are neglected. Indeed, 
\begin{equation*}
\left\langle A^{i}(k)A^{j}(-k)\right\rangle _{0}\rightarrow \frac{i\delta
^{ij}}{E^{2}-\mathbf{k}^{2}-m_{\gamma }^{2}}+\frac{i(1-\lambda )k^{i}k^{j}}{%
(E^{2}-\mathbf{k}^{2}-m_{\gamma }^{2})(\lambda E^{2}-\mathbf{k}^{2}-\lambda
m_{\gamma }^{2})}.
\end{equation*}%
To get rid of the denominators $\mathbf{k}^{2}$ and the coinciding
thresholds at the same time, we can make the same $\eta $ deformation in
both terms of each propagator $\left\langle A^{i}(k)A^{j}(-k)\right\rangle
_{0}$. In the end, the quantization (\ref{specialone}) ensures that the
ultraviolet divergences are local and the counterterms obey the usual rules
of power counting, so the proof of renormalizability reveals no surprises.
Moreover, the counterterms are polynomial in the masses (and $\eta $).

At $m_{0}=m_{\gamma }=\bar{m}=0$, we have a renormalization constant $Z_{g}$
for the gauge coupling $g$ and wave-function renormalization constants $%
Z_{0} $, $Z_{\gamma }$ and $\bar{Z}$ for $A_{0}$, $\mathbf{A}$ and $\bar{C}$-%
$C$, respectively. At nonvanishing gauge masses, the renormalized Lagrangian
coincides with the one at $m_{0}=m_{\gamma }=\bar{m}=0$ plus the
counterterms 
\begin{equation*}
\Delta \mathcal{L}_{m_{\gamma }}=\frac{\Delta m_{0}^{2}}{2}A_{0}^{2}-\frac{%
\Delta m_{\gamma }^{2}}{2}\mathbf{A}^{2}-\Delta \bar{m}^{2}\bar{C}C,
\end{equation*}
where $\Delta m_{0}^{2}$, $\Delta m_{\gamma }^{2}$ and $\Delta \bar{m}^{2}$
are divergent constants.

In the limit $\lambda \rightarrow 1$ we can choose Lorentz invariant mass
terms ($m_{0}=m_{\gamma }$). In that case, the action (\ref{lmg}) is Lorentz
invariant, as well as its renormalization, so $Z_{0}=Z_{\gamma }$ and $%
\Delta m_{0}^{2}=\Delta m_{\gamma }^{2}$. However, the finite parts of the
amplitudes are not exactly Lorentz invariant, because different quantization
prescriptions are used for the physical and unphysical poles of the
propagators (which are distinguished from one another in a non-Lorentz
invariant way). The Lorentz violations appear starting from the imaginary
parts of the one-loop diagrams, above the fake thresholds. Lorentz symmetry
is recovered in the limit of vanishing gauge masses (see below).

\subsubsection*{Unitarity}

The theory is perturbatively unitary, even at nonvanishing gauge masses,
because both the Feynman prescription and the fakeon prescription are
manifestly consistent with unitarity \cite{fakeons}. The loop integrals are
evaluated at $\eta \neq 0$ as explained above. It is crucial to observe that
the cutting equations, which are identities that can be written down for
every diagram separately, hold for arbitrary $\eta \neq 0$. Then, they still
hold in the limit $\eta \rightarrow 0$, which proves the optical theorem.

\subsubsection*{Gauge invariance and gauge independence}

The next task is to prove that gauge invariance is recovered in the limit of
vanishing gauge masses. Gauge invariance is expressed by means of the
Slavnov-Taylor-Ward-Takahashi (STWT) identities \cite{WTST}, which establish
relations among (off-shell) amplitudes and loop diagrams. Such identities
can be collected into the Zinn-Justin equation \cite{ZJ}, also-called master
equation, which can be written as $(\Gamma ,\Gamma )=0$ (assuming that we
use the dimensional regularization), where $\Gamma $ is the generating
functional of the one-particle irreducible diagrams and $(.,.)$ denotes the
Batalin-Vilkovisky antiparentheses \cite{BV}.

In the absence of fakeons, the limit of vanishing gauge masses is smooth
off-shell, so we\ can set them directly to zero in the integrands of the
loop diagrams. When fakeons are present we have to be more careful, because
we need to work at $\eta \neq 0$ to avoid the coinciding thresholds, which
in turn requires nonvanishing gauge masses.

Recall that the STWT identities stem from simple, polynomial relations among
the Feynman rules. The famous QED\ Ward identity, for example, follows from%
\begin{equation}
\gamma ^{\mu }k_{\mu }-\left[ \gamma ^{\mu }(p+k)_{\mu }-m\right] +\gamma
^{\mu }p_{\mu }-m=0.  \label{poli}
\end{equation}%
In other words, even the more complicated STWT identity can be phrased as
the loop integral of a rational function $r(q)$ that factorizes a polynomial
that vanished identically, such as the left-hand side of (\ref{poli}), where 
$q$ denotes all the momenta involved. At $\eta \neq 0$, $m_{g}\neq 0$ (where 
$m_{g}$ denotes the gauge masses), $r(q)=0$ turns into a corrected algebraic
relation of the form 
\begin{equation}
r(q,\eta ,m_{g})=\eta r^{\prime }(q,\eta ,m_{g})+m_{g}^{2}r^{\prime \prime
}(q,\eta ,m_{g}),  \label{rp}
\end{equation}%
where both sides are rational functions, $r(q,0,0)=r(q)$ and $r^{\prime
}(q,\eta ,m_{g})$ and $r^{\prime \prime }(q,\eta ,m_{g})$ are regular for $%
\eta \rightarrow 0$, $m_{g}\rightarrow 0$. For instance, in the case of (\ref%
{poli}), if we deform the masses we obtain 
\begin{equation}
\gamma ^{\mu }k_{\mu }-\left[ \gamma ^{\mu }(p+k)_{\mu }-m_{1}\right] +\left[
\gamma ^{\mu }p_{\mu }-m_{2}\right] =m_{1}-m_{2}\equiv \eta ,  \label{qe}
\end{equation}%
where the left-hand side stands for $r(q,\eta )$ and $r^{\prime }=1$, $%
r^{\prime \prime }=0$.

When we integrate on the loop momenta, both sides of (\ref{rp}) have no
coinciding thresholds. We start again from the Euclidean region, then move
to the other regions by taking the average continuation where necessary and
finally take the limit $\eta \rightarrow 0$ of coinciding thresholds. The
first term on the right hand side of (\ref{rp}) disappears in the limit. The
second term describes the violation of gauge invariance at nonvanishing
gauge masses $m_{g}$ and disappears in the limit $m_{g}\rightarrow 0$. This
proves the STWT identities.

Normally, when we manipulate identities like (\ref{rp}), $r(q,\eta ,m_{g})$
is a sum of terms that end up being part of different diagrams, which are
calculated separately. Thus, it is important to overcome the thresholds
consistently in all of them. The prescription formulated\ so far ensures
this, by\ treating all the $\lambda $-dependent thresholds by means of the
average continuation and all the $\lambda $-independent thresholds by means
of the analytic continuation.

To show that gauge independence is also recovered in the limit of vanishing
gauge masses, we can argue similarly. Indeed, gauge independence also stems
from simple polynomial identities obeyed by the Feynman rules.

\bigskip

Lorentz invariance is broken by the quantization prescription (\ref%
{specialone}). However, it is recovered in the limit of vanishing gauge
masses. Precisely, once gauge invariance and gauge independence are
restored, the Lorentz violation is confined to the gauge-trivial sector of
the theory, which does not affect the physical quantities.

On the other hand, when the gauge masses are nonvanishing, the physical
quantities are not Lorentz invariant. As stressed above, the Lorentz
violation can be \textquotedblleft minimized\textquotedblright\ by taking
the limit $\lambda \rightarrow 1$ (which can be done only at the end of the
calculations, since the $\lambda $ dependence is crucial to distinguish the
fake thresholds from the physical ones) and choosing $m_{0}=m_{\gamma }$.

In conclusion, the quantization formulated here is manifestly unitary for
arbitrary gauge masses. It is gauge and Lorentz invariant in the limit of
vanishing gauge masses.

Observe that the set of physical degrees of freedom is always the same, at
vanishing and nonvanishing gauge masses, since the fakeons are always
projected away from the physical spectrum. It is evident that the proof of
unitarity we have just provided is much more economic than any other proof
given so far \cite{thooft,unitarityc}.

Normally, an explicit breaking of gauge invariance, such as the one due to
the gauge masses, is expected to break unitarity, by making unphysical
degrees of freedom propagate. This does not happen, if we quantize the
would-be unphysical degrees of freedom as fakeons. A byproduct of our
construction is that we can build manifestly unitary, local, renormalizable
theories of massive gauge fields.

\subsubsection*{Higher dimensions}

In higher dimensions the theory (\ref{lmg}) is nonrenormalizable. The
gauge-fixing procedure, the propagators and the quantization prescriptions (%
\ref{specialone}) are the same. The only part that changes is the set of
counterterms, which are infinitely many.

At nonvanishing gauge masses, Lorentz violating counterterms appear in both
the physical and gauge sectors, multiplied by the gauge masses. When we
include them, we basically have a theory of scalar fields and space vector
fields. The quadratic terms can be resummed into \textquotedblleft
dressed\textquotedblright\ propagators 
\begin{equation}
\langle A_{\mu }A_{\nu }\rangle _{\mathrm{dressed}}=\langle A_{\mu }A_{\nu
}\rangle _{0}+\langle A_{\mu }A_{\rho }\rangle _{0}V^{\rho \sigma }\langle
A_{\sigma }A_{\nu }\rangle _{0}+\cdots ,  \label{resumma}
\end{equation}%
where $V^{\rho \sigma }$ is local and collects the quadratic terms of higher
dimensions turned on by renormalization. What is important is that the
dressed propagators still have the properties we need to prove unitarity
along the guidelines explained above. In particular, using the arguments of
ref. \cite{abse} we can remove all the higher time derivatives from the
quadratic action (and so $V^{\rho \sigma }$) by means of field
redefinitions: this ensures that the resummation (\ref{resumma}) generates
no new poles. The physical poles remain $\lambda $ independent and the
unphysical poles remain $\lambda $ dependent, so we can distinguish the
physical thresholds from the fake ones inside the loop diagrams and treat
them accordingly. In the end, the proof of unitarity works as above. In the
limit of vanishing gauge masses, the counterterms are gauge and Lorentz
invariant in the physical sector and rotationally invariant in the
gauge-trivial sector.

By means of the fakeon quantization prescription, it is possible to build
local, unitary, strictly renormalizable Yang-Mills theories in arbitrary
higher spacetime dimensions $d\geqslant 6$ \cite{correspondence}. Their
interim classical actions read 
\begin{equation}
S_{\mathrm{YM}}^{d}=-\frac{1}{4}\int \mathrm{d}^{d}x\hspace{0.01in}F_{\mu
\nu }^{a}P(D^{2})F^{a\mu \nu }+\mathcal{O}(F^{3})\mathrm{,}  \label{YMD}
\end{equation}%
where $D$ is the covariant derivative, $P(x)$ is a real polynomial of degree 
$(d-4)/2$ in $x$ such that $P(0)>0$, while $\mathcal{O}(F^{3})$ are the
Lagrangian terms that have dimensions smaller than or equal to $d$ and are
built with at least three field strengths and their covariant derivatives.
The quadratic terms can always be reduced to the form (\ref{YMD}) by means
of Bianchi identities and partial integrations. The coefficients of the
polynomial $P$ must be such that the poles of $1/P$ are massive and the
squared masses have nonnegative real parts.

The special gauge can be built by choosing the gauge-fixed Lagrangian 
\begin{eqnarray}
\mathcal{L}_{\mathrm{gf}} &=&-\frac{1}{4}F_{\mu \nu }P(D^{2})F^{\mu \nu }+%
\mathcal{O}(F^{3})-\frac{1}{2\lambda }\mathcal{G}(A)P(\partial ^{2})\mathcal{%
G}(A)-\bar{C}P(\partial ^{2})\mathcal{G}(DC)  \notag \\
&&+\frac{1}{2}A^{\mu }\left( m_{0}^{2}\delta _{\mu 0}\delta _{0\nu
}-m_{\gamma }^{2}\delta _{\mu i}\delta _{i\nu }\right) P(\partial
^{2})A^{\nu }-\bar{m}^{2}\bar{C}P(\partial ^{2})C.  \label{YMDgf}
\end{eqnarray}%
We have chosen convenient \textquotedblleft mass terms\textquotedblright ,
to simplify the propagators, which then coincide with the ones of (\ref%
{specialone}), multiplied by $1/P(-k^{2})$. The quantization prescription
follows from the replacements 
\begin{eqnarray*}
\frac{1}{(\lambda E^{2}-\mathbf{k}^{2}-m^{2})P(-k^{2})} &\rightarrow &\left. 
\frac{1}{(\lambda E^{2}-\mathbf{k}^{2}-m^{2})P(-k^{2})}\right\vert _{\text{f}%
}, \\
\frac{1}{(k^{2}-m^{2})P(-k^{2})} &\rightarrow &\frac{1}{(k^{2}-m^{2}+i%
\epsilon )P(-m^{2})}-\left. \frac{P(-k^{2})-P(-m^{2})}{%
P(-k^{2})(k^{2}-m^{2})P(-m^{2})}\right\vert _{\ast },
\end{eqnarray*}%
where $m$ is $m_{0}$, $m_{\gamma }$ or $\bar{m}$, depending on the case. The
star in the second line means that the poles with negative or complex
residues, as well as those with positive residues but complex masses, must
be quantized as fakeons. Instead, the poles with positive residues and
nonvanishing real masses can be quantized either as fakeons or physical
particles.

Renormalization generates mass terms of lower dimensionalities, but we do
not need to include them at the tree level, since they are going to
disappear when we take the gauge masses to zero. The proof of unitarity
proceeds as above, as well as the recovery of gauge invariance and Lorentz
invariance at vanishing gauge masses.

\section{Quantum gravity: low-energy theory}

\label{qg1}

\setcounter{equation}{0}

In this section and the next ones we generalize the proof to quantum gravity
in arbitrary dimensions $d>3$. We start from the low-energy
nonrenormalizable theory at vanishing cosmological constant. In the next
section we formulate the theory of scattering at $\Lambda \neq 0$ and in
section \ref{qg2} we generalize the results to ultraviolet complete theories.

The gauge-fixed Hilbert-Einstein Lagrangian is 
\begin{equation}
\mathcal{L}_{\mathrm{gf}}=-\frac{1}{2\kappa ^{d-2}}\sqrt{|g|}R+\frac{1}{%
4\lambda _{1}\kappa ^{d-2}}\mathcal{G}_{0}^{2}(g)-\frac{1}{4\lambda
_{2}\kappa ^{d-2}}\mathcal{G}_{i}^{2}(g)+\bar{C}_{0}\mathcal{G}_{0}(%
\overline{DC})-\bar{C}_{i}\mathcal{G}_{i}(\overline{DC}),  \label{lgf}
\end{equation}
where $\mathcal{G}_{0}(g)$ and $\mathcal{G}_{i}(g)$ are the gauge-fixing
functions, assumed to be linear in the metric $g_{\mu \nu }$, while $C_{\mu
} $ and $\bar{C}_{\mu }$ are the Faddeev-Popov ghosts and antighosts,
respectively, and $\overline{DC}$ stands for $D_{\mu }C_{\nu }+D_{\nu
}C_{\mu }$, $D_{\mu }$ denoting the covariant derivative. The constant $%
\kappa $ is chosen to have dimension $-1$ in units of mass for every $d$.

The special gauge is obtained by choosing \cite{unitarityc} 
\begin{equation}
\mathcal{G}_{0}(g)=\frac{\lambda }{2}\partial _{0}g_{00}+\frac{1}{2}\partial
_{0}g_{ii}-\partial _{i}g_{0i},\qquad \mathcal{G}_{i}(g)=-\lambda
_{1}\partial _{j}g_{ij}+\frac{1}{2}\left( 2\lambda _{1}-1\right) \partial
_{i}g_{jj}+\lambda \partial _{0}g_{0i}-\frac{\lambda }{2}\partial _{i}g_{00},
\label{specialtwo}
\end{equation}%
with 
\begin{equation*}
\lambda _{1}=\frac{\lambda (d-3)+d-1}{2(d-2)},\qquad \lambda _{2}=\lambda
\lambda _{1}.
\end{equation*}

We expand around flat space by writing $g_{\mu \nu }=\eta _{\mu \nu
}+2\kappa ^{(d/2)-1}h_{\mu \nu }$. With the prescriptions 
\begin{equation*}
\bar{P}_{\mathrm{phys}}=\frac{1}{E^{2}-\mathbf{k}^{2}+i\epsilon },\qquad 
\bar{P}_{\text{f}}=\left. \frac{1}{\lambda E^{2}-\mathbf{k}^{2}}\right\vert
_{\text{f}},\qquad \bar{P}_{\text{f}}^{\prime }=\left. \frac{1}{\lambda
E^{2}-\lambda _{1}\mathbf{k}^{2}}\right\vert _{\text{f}},
\end{equation*}%
we find the ghost propagators 
\begin{equation}
\langle C^{0}\bar{C}^{0}\rangle _{0}=-i\bar{P}_{\text{f}},\qquad \langle
C^{0}\bar{C}^{i}\rangle _{0}=\langle C^{i}\bar{C}^{0}\rangle _{0}=0,\qquad
\langle C^{i}\bar{C}^{j}\rangle _{0}=i\bar{P}_{\text{f}}^{\prime }\Pi ^{ij}+i%
\bar{P}_{\text{f}}\frac{k^{i}k^{j}}{\mathbf{k}^{2}},  \label{pgh}
\end{equation}%
and the $h_{\mu \nu }$ propagators 
\begin{eqnarray}
\langle h_{00}h_{00}\rangle _{0} &=&\frac{d-3}{d-2}i\bar{P}_{\text{f}%
},\qquad \langle h_{00}h_{ij}\rangle _{0}=\frac{i\delta _{ij}\bar{P}_{\text{f%
}}}{d-2},  \notag \\
\langle h_{0i}h_{0j}\rangle _{0} &=&-\frac{i\lambda _{1}}{2}\left( \bar{P}_{%
\text{f}}^{\prime }\Pi _{ij}+\bar{P}_{\text{f}}\frac{k_{i}k_{j}}{\mathbf{k}%
^{2}}\right) ,\qquad \langle h_{00}h_{0i}\rangle _{0}=\langle
h_{0i}h_{jk}\rangle _{0}=0,  \label{propag} \\
\langle h_{ij}h_{mn}\rangle _{0} &=&\frac{i\bar{P}_{\mathrm{phys}}}{2}\left(
\Pi _{im}\Pi _{jn}+\Pi _{in}\Pi _{jm}-\frac{2}{d-2}\Pi _{ij}\Pi _{mn}\right)
-\frac{\lambda }{\mathbf{k}^{2}}\frac{i\bar{P}_{\text{f}}}{d-2}\left( \Pi
_{ij}k_{m}k_{n}+k_{i}k_{j}\Pi _{mn}\right)  \notag \\
&&\!\!\!\!\!{+\frac{\lambda i\bar{P}_{\text{f}}^{\prime }}{2\mathbf{k}^{2}}%
\left( \Pi _{im}k_{j}k_{n}+\Pi _{in}k_{j}k_{m}+\Pi _{jm}k_{i}k_{n}+\Pi
_{jn}k_{i}k_{m}\right) +\lambda i\bar{P}_{\text{f}}\frac{d-3}{d-2}\frac{%
k_{i}k_{j}k_{m}k_{n}}{(\mathbf{k}^{2})^{2}}.}  \notag
\end{eqnarray}

As in the case of gauge theories, the denominators proportional to $\mathbf{k%
}^{2}$ and $(\mathbf{k}^{2})^{2}$ cancel out, if the quantization
prescriptions are ignored. This ensures that the locality of counterterms
works as usual, since the ultraviolet divergences do not depend on the
prescriptions.

To have control on the on-shell infrared divergences, we add the most
general mass terms that are invariant under rotations, 
\begin{equation}
\Delta \mathcal{L}_{m}=-\frac{m_{1}^{2}}{4}h_{00}^{2}-\frac{m_{2}^{2}}{2}%
h_{ij}^{2}+m_{3}^{2}h_{0i}^{2}+\frac{m_{4}^{2}}{4}h_{ii}h_{jj}-\frac{%
m_{5}^{2}}{2}h_{00}h_{ii}+\frac{\bar{m}_{1}^{2}}{2}\bar{C}^{0}C^{0}-\frac{%
\bar{m}_{2}^{2}}{2}\bar{C}^{i}C^{i}.  \label{deltalm}
\end{equation}%
The coefficients are labeled so that when all the gauge masses $m_{a}$, $%
a=1,2,3,4$, are equal to $m$ and the ghost masses $\bar{m}_{b}$, $b=1,2$,
are equal to $\bar{m}$, we obtain the Lorentz invariant combination 
\begin{equation}
\Delta \mathcal{L}_{m}=-\frac{m^{2}}{2}\left( h_{\mu \nu }h^{\mu \nu }-\frac{%
1}{2}h^{2}\right) +\frac{\bar{m}^{2}}{2}\bar{C}^{\mu }C_{\mu }.
\label{dlminv}
\end{equation}%
As far as the graviton propagator is concerned, the cosmological constant
can be seen as a correction to $m^{2}$ (see next section).

The propagators for the most general mass terms (\ref{deltalm}) are rather
involved. We just report that in both cases (\ref{deltalm}) and (\ref{dlminv}%
) they have no simple poles. Moreover, the poles have squared masses with
positive real parts if 
\begin{equation*}
m_{1}^{2}>0,\qquad m_{2}^{2}>0,\qquad m_{3}^{2}>0,\qquad
(d-1)m_{4}^{2}>2m_{2}^{2},\qquad m_{5}^{2}>0
\end{equation*}%
(in addition to $\lambda >0$, $d>3$). From now on, we assume that such
inequalities hold. As in (\ref{propag}), the unphysical poles are $\lambda $
dependent and the physical poles are $\lambda $ independent.

Without making involved calculations, the $\lambda $ dependence can be
studied as follows. The massive propagators 
\begin{equation}
\langle h_{\mu \nu }h_{\rho \sigma }\rangle _{m}=\langle h_{\mu \nu }h_{\rho
\sigma }\rangle _{0}+\langle h_{\mu \nu }h_{\alpha \beta }\rangle
_{0}V_{m}^{\alpha \beta \gamma \delta }\langle h_{\gamma \delta }h_{\rho
\sigma }\rangle _{0}+\cdots ,  \label{masspr}
\end{equation}%
can be obtained by resumming the corrections due to the two-leg vertices $%
V_{m}^{\alpha \beta \gamma \delta }$ provided by $\Delta \mathcal{L}_{m}$.
The projector 
\begin{equation*}
\frac{1}{2}\left( \Pi _{im}\Pi _{jn}+\Pi _{in}\Pi _{jm}-\frac{2}{d-2}\Pi
_{ij}\Pi _{mn}\right) ,
\end{equation*}%
which multiplies the physical pole in (\ref{propag}), is orthogonal to every
term we may build for $\langle h_{\mu \nu }h_{\rho \sigma }\rangle _{m}$,
apart from the identity $(\delta ^{im}\delta ^{jn}+\delta ^{in}\delta
^{jm})/2$. Moreover, it cannot be generated by multiplying terms that do not
contain the identity. For this reason, when we perform the resummation (\ref%
{masspr}), the $\lambda $-dependent poles do not affect the physical pole,
and vice versa. Note that new poles may appear in the resummation, because
some invariants on the right-hand sides of (\ref{propag}) are missing. By
the arguments just given, such new poles are necessarily $\lambda $
dependent and must be quantized as fakeons.

The theory is nonrenormalizable. At vanishing gauge masses, we must include
all the local, generally covariant terms that can be built with at least
three Weyl tensors and their covariant derivatives \cite{newQG}, multiplied
by independent parameters. Then, the divergent parts of the Feynman diagrams
are subtracted by means of redefinitions of the parameters and the fields.
Note that, by power counting, the cosmological term is not generated, since
the theory contains no parameters of positive dimensions in units of mass.
When the gauge masses are nonvanishing, extra counterterms proportional to
the squared gauge masses must be added. They do not need to be general
covariant, but just invariant under space rotations.

In the evaluation of the loop diagrams, the $\lambda $ dependent thresholds
of the fake processes can be distinguished from the thresholds of the
physical processes, which are $\lambda $ independent. As in case of gauge
theories, this allows us to circumvent the former by means of the average
continuation and the latter by means of the Feynman prescription, thereby
proving perturbative unitarity.

General covariance is recovered in the limit of vanishing gauge masses. When 
$\lambda =1$ we can choose the $\Delta \mathcal{L}_{m}$ of formula (\ref%
{dlminv}) to have a Lorentz invariant renormalization. The finite parts of
the amplitudes, however, are not Lorentz and general covariant. They become
so only when the gauge masses are sent to zero.

So far, we have set the cosmological constant $\Lambda $ to zero, which is
consistent only in special, unrealistic models. The problem of defining the
theory of scattering in the presence of a cosmological constant must be
discussed apart.

\section{Theory of scattering in the presence of a cosmological constant}

\label{qed}

\setcounter{equation}{0}

In this section we formulate the theory of scattering in the presence of a
small, but nonvanishing cosmological constant $\Lambda $. By expanding
around flat space, we obtain scattering amplitudes that satisfy perturbative
unitarity up to corrections due to $\Lambda $. Such corrections are
negligible for all practical purposes. In the academic case they were non
negligible, our analysis provides well-defined cutting equations, which
however do not have a clear physical interpretation in the context of a
theory of scattering. For definiteness, we assume to work in four
dimensions, but the arguments work in arbitrary dimensions $d>3$.

\bigskip

Let us first address the main aspects of the problem we have to deal with.
In some models, the cosmological constant $\Lambda $ can be turned off
consistently, since the $\Lambda $ beta function vanishes when $\Lambda $
vanishes. The simplest example is pure gravity, whose Lagrangian is the sum
of the Hilbert term, plus the counterterms built with at least three Weyl
tensors and their covariant derivatives \cite{newQG}. There, power counting
ensures that the cosmological constant is not turned on by renormalization,
because the theory contains no parameters of positive dimensions in units of
mass. In the realm of ultraviolet complete theories, $\Lambda $ can be
consistently switched off in super-renormalizable models with more higher
derivatives \cite{LWgrav}. Both types of models, however, are not realistic,
since $\Lambda $ is turned on by renormalization as soon as massive or
self-interacting matter fields are included.

Thus, it is compulsory to study the case $\Lambda \neq 0$ in detail.
However, a consistent theory of scattering is available only in flat space
and might not even exist at $\Lambda \neq 0$ \cite{adsscat}, where we cannot
talk about asymptotic states and scattering amplitudes in a strict sense. At
the same time, flat space is not a solution of the classical field equations
(in the absence of matter) at $\Lambda \neq 0$, and the perturbative
expansion around nonflat backgrounds is extremely inconvenient.

When we study scattering experiments for our laboratories, we do not care
whether the universe has a cosmological constant or not. We just expand
around flat space and move on. Since the value of $\Lambda $ is very small
in nature, we expect that its effects are negligible for all practical
purposes. Thus, in the presence of a cosmological constant it should be
possible to formulate a theory of scattering that makes physical sense up to
the corrections due to $\Lambda $. Put it differently, we demand that the
theory be \textquotedblleft as unitary as it can be\textquotedblright\ at $%
\Lambda \neq 0$.

At the classical level, it is obvious that $\Lambda $ can be treated
perturbatively and neglected for most purposes. It is not obvious that we
can do the same in quantum field theory. Indeed, often quantities that are
classically negligible become important due to quantum effects. For example,
the axial anomalies and the renormalization group flow are originated by
conflicts between classically negligible quantities and ultraviolet
divergences. In the case of the cosmological constant, a possible source of
conflict is provided by the infrared divergences.

For these reasons, we need to investigate the matter carefully. We insist on
expanding around flat space and our results show that, in the end, this is
the right choice.

\bigskip

When we expand around flat space, the cosmological term 
\begin{equation*}
-\frac{\Lambda }{\kappa ^{2}}\int \mathrm{d}^{4}x\sqrt{-g}
\end{equation*}
generates

$i$) tadpole (one-leg) vertices, which allow us to build infinitely many
connected diagrams of the same order;

$ii$) quadratic terms (two-leg vertices), which can be resummed to give the
graviton a sort of \textquotedblleft mass\textquotedblright ;

$iii$)\ super-renormalizable vertices, which cause the appearance of
(off-shell) infrared divergences in loop diagrams (for $\Lambda $ small).

Specifically, point ($ii$) leads to a graviton propagator that reads, in the
De Donder gauge, 
\begin{equation}
\mathcal{\langle }h_{\mu \nu }(k)\hspace{0.01in}h(-k)\mathcal{\rangle }_{0}=%
\frac{i}{2}\frac{\eta _{\mu \rho }\eta _{\nu \sigma }+\eta _{\mu \sigma
}\eta _{\nu \rho }-\eta _{\mu \nu }\eta _{\rho \sigma }}{k^{2}+2\Lambda
+i\epsilon }.  \label{tac}
\end{equation}%
This may look okay when $\Lambda <0$, but is tachyonic when $\Lambda >0$.

Let us see how to overcome the problems just listed one by one. The problems
($i$) appear because the propagator (\ref{tac}) is proportional to $%
1/\Lambda $ in the infrared limit $k\rightarrow 0$ and the tadpole vertices
are of order $\Lambda $. To better illustrate the issue, consider the
Lagrangian 
\begin{equation}
\mathcal{L}_{\Lambda }=-\frac{\Lambda }{\kappa ^{2}}\sqrt{-g}.
\label{lsimple}
\end{equation}%
Expanding the metric around flat space, by writing $g_{\mu \nu }=\eta _{\mu
\nu }+2\kappa h_{\mu \nu }$, the Legendre transform $\mathcal{F}$ of $%
\mathcal{L}_{\Lambda }$ with respect to $h_{\mu \nu }$ gives 
\begin{equation*}
\mathcal{F}(J)=\mathcal{L}_{\Lambda }-J^{\mu \nu }h_{\mu \nu }=\frac{1}{%
\Lambda }\sqrt{-\det J^{\mu \nu }}+\frac{1}{2\kappa }J^{\mu \nu }\eta _{\mu
\nu },
\end{equation*}%
where 
\begin{equation}
J^{\mu \nu }=\frac{\partial \mathcal{L}_{\Lambda }}{\partial h_{\mu \nu }}=-%
\frac{\Lambda }{\kappa }\sqrt{-g}g^{\mu \nu }.  \label{j}
\end{equation}%
Now, if $\mathcal{L}_{\Lambda }$ is viewed as a classical Lagrangian, it
provides well-defined Feynman rules. Moreover, since the $\mathcal{L}%
_{\Lambda }$ quadratic part is non dynamic, all loop diagram vanish (in
dimensional regularization), so $\mathcal{L}_{\Lambda }$ coincides with the
generating functional of the one-particle irreducible diagrams. Then, $%
\mathcal{F}$ should be the generating functional of the corresponding
connected diagrams, which would be the derivatives of $\mathcal{F}$ with
respect to $J^{\mu \nu }$, evaluated at $J^{\mu \nu }=0$. However, such
derivatives are generically singular. The reason is that, by means of
one-leg vertices, it is possible to build infinitely many connected tree
diagrams of the same order, with the same set of external legs.

The point is that setting $J^{\mu \nu }=0$ is not the right thing to do,
since it does not correspond to flat space. Actually, (\ref{j}) shows that
the condition $J^{\mu \nu }=0$ implies $\sqrt{-g}g^{\mu \nu }=0$, which
leads to singularities. Since we insist on expanding around flat space, we
should impose conditions that correspond to flat space both before and after
the Legendre transform. The right source is then 
\begin{equation}
j^{\mu \nu }=J^{\mu \nu }+\frac{\Lambda }{\kappa }\eta ^{\mu \nu }
\label{source}
\end{equation}%
and the connected diagrams are the derivatives of $\mathcal{F}$ with respect
to $j^{\mu \nu }$, calculated at $j^{\mu \nu }=0$. As a check, it is
straightforward to verify that the propagator, obtained by inverting the
second derivative of $\mathcal{L}_{\Lambda }$ with respect to $h_{\mu \nu }$%
, calculated at $h_{\mu \nu }=0$, is equal to the second derivative of $%
\mathcal{F}$ with respect to $j^{\mu \nu }$, calculated at $j^{\mu \nu }=0$.

In conclusion, the definition (\ref{source}) removes problem ($i$). In
practice, it removes the connected diagrams that contain legs attached to
tadpole vertices.

The problem ($ii$) of the tachyonic propagator (\ref{tac}) for $\Lambda >0$
can be overcome by introducing a mass term for the graviton. Let us consider
the Lagrangian 
\begin{equation*}
\mathcal{L}_{\Lambda }^{\prime }=-\frac{\Lambda }{\kappa ^{2}}\sqrt{-g}-%
\frac{1}{2}h_{\mu \nu }(\square +m_{g}^{2})h^{\mu \nu }+\frac{1}{4}h(\square
+m_{g}^{2})h.
\end{equation*}
and treat $\Lambda $ perturbatively with respect to $m_{g}^{2}$. The
graviton propagator in the De Donder gauge, 
\begin{equation}
\mathcal{\langle }h_{\mu \nu }(k)\hspace{0.01in}h(-k)\mathcal{\rangle }_{0}=%
\frac{i}{2}\frac{\eta _{\mu \rho }\eta _{\nu \sigma }+\eta _{\mu \sigma
}\eta _{\nu \rho }-\eta _{\mu \nu }\eta _{\rho \sigma }}{k^{2}-m_{g}^{2}+2%
\Lambda +i\epsilon },  \label{tempo}
\end{equation}
is not tachyonic as long as $m_{g}^{2}>2\Lambda $. If this inequality is
satisfied, we face no further obstruction to derive the cutting equations.

One might object that the graviton is massless in nature, so at the end $%
m_{g}$ should tend to zero. Hence, it does not seem to make sense to take $%
m_{g}^{2}>2\Lambda $. However, we stress again that our purpose is not to
formulate a theory of scattering at nonzero $\Lambda $ in a strict sense,
which is likely impossible. We just want to formulate a theory of scattering
at $\Lambda \neq 0$ that is meaningful up to corrections due to $\Lambda $
itself. If $m_{g}^{2}$ is sufficiently small, larger than $2\Lambda $ and
such that $|m_{g}^{2}-2\Lambda |\sim |\Lambda |$ (for definiteness, it may
be useful to assume $m_{g}^{2}\sim 3|\Lambda |$), we can keep it
nonvanishing as well, since its corrections are not so different from the
ones due to $\Lambda $, and whenever the latter are negligible, so are the
former. In this sense, the solution (\ref{tempo}) removes problem ($ii$). 
\begin{figure}[t]
\begin{center}
\includegraphics[width=8truecm]{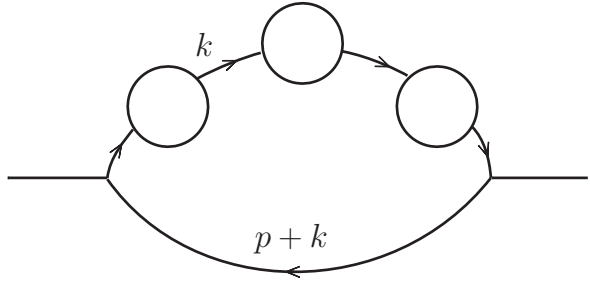}
\end{center}
\caption{Multibubble diagrams}
\label{multibubble}
\end{figure}

As said, we have well-defined cutting equations whenever $m_{g}^{2}>2\Lambda 
$. However, the tools that allow us to prove unitarity from the cutting
equations, which are asymptotic states, scattering amplitudes and reduction
formulas, are well-defined only at $\Lambda =0$. Thus, we can at most prove
unitarity up to the corrections due to $\Lambda $.

To ensure that the effects of $\Lambda $ are indeed small enough to be
negligible, we study the behavior of the physical quantities when both $%
m_{g}^{2}$ and $\Lambda $ tend to zero at the same time and $%
|m_{g}^{2}-2\Lambda |\sim |\Lambda |$. We avoid problems with tachyonic
poles by keeping $m_{g}^{2}>2\Lambda $, or analogous variants of such an
inequality [which apply when $\lambda \neq 1$ and (\ref{deltalm}) is used].
We know that in realistic models the limit $|m_{g}^{2}-2\Lambda |\sim
|\Lambda |\rightarrow 0$ cannot be pushed to the very end, since it is not
consistent to turn $\Lambda $ identically off. Thus, we stop short of doing
that, which is anyway enough to estimate the corrections due to $\Lambda
\neq 0$.

The main source of worry is problem ($iii$), since certain off-shell
infrared divergences are generated inside the loop diagrams in the limit.
Specifically, the divergences appear when propagators with the same loop
momentum $p$ are raised to high powers, which happens in multibubble
diagrams like the one of fig. \ref{multibubble}.

Let 
\begin{equation*}
B(p^{2})\sim \sum_{i}C_{i}\int_{0}^{1}\mathrm{d}x\ln \left[
m_{i}^{2}-i\epsilon -p^{2}x(1-x)\right]
\end{equation*}
denote the bubble diagram with external graviton legs and external momentum $%
p$, where $C_{i}$ are factors due to the vertices. For simplicity, we are
assuming that the fields circulating in the loop have the same masses $m_{i}$
and the vertices contributing to $B(p^{2})$ are non derivative, since our
estimates will not depend on such assumptions.

If $P$ denotes the graviton propagator, the loop integral of the multibubble
diagrams have the following infrared behaviors: 
\begin{eqnarray}
\int_{\mathrm{IR}}PB\cdots BP &\sim &\int_{\mathrm{IR}}\frac{\mathrm{d}^{4}p%
}{(2\pi )^{4}}\frac{B(p^{2})^{n}}{(p^{2}-m_{g}^{2}+2\Lambda +i\epsilon
)^{n+1}}\sim \frac{B(0)^{n}}{(m_{g}^{2}-2\Lambda -i\epsilon )^{n-1}}\qquad 
\mathrm{for}\text{ }n>1,  \notag \\
\int_{\mathrm{IR}}PBP &\sim &B(0)\ln (m_{g}^{2}-2\Lambda -i\epsilon ).
\label{IR}
\end{eqnarray}%
Again, we assume that the vertices are non derivative, since extra powers of 
$p$ carried by them can only improve the infrared behaviors. The $i\epsilon $
prescription is kept to emphasize that we have a well-defined way to cross $%
m_{g}^{2}=2\Lambda $, so it does not really matter whether $\Lambda $ is
positive or negative, as long as $|m_{g}^{2}-2\Lambda |\sim |\Lambda |$.

To estimate the corrections due to $m_{g}$ and $\Lambda $, when they are
small, it is convenient to resum the bubble diagram into the corrected
propagator 
\begin{equation}
\frac{1}{p^{2}-M^{2}(p^{2})+i\epsilon },  \label{corp}
\end{equation}%
where 
\begin{equation}
M^{2}(p^{2})\equiv m_{g}^{2}-2\Lambda +B(p^{2}).  \label{runL}
\end{equation}%
It is easy to show that the $M$ corrections to a loop diagram calculated
with the propagators (\ref{corp}) are of order $M^{2}\ln M^{2}$ for $M$
small [the logarithm being due to integrals $\sim \mathrm{d}%
^{4}p/(p^{2})^{2} $ originated by the small-$M$ expansion]. Now, in the
limit we want to study, the absolute value of $M^{2}$ is of order $|\Lambda
_{R}|$, where $\Lambda _{R}$ is the running cosmological constant, so in the
end the corrections due to $\Lambda $ are 
\begin{equation}
|\Lambda _{R}|\ln |\Lambda _{R}|,  \label{lr}
\end{equation}%
to be divided by an energy squared.

The radiative corrections $\Delta \Lambda $ of $\Lambda _{R}=\Lambda +\Delta
\Lambda $, due to the one-loop diagrams, are 
\begin{equation}
\Delta \Lambda \sim \frac{m^{4}}{M_{\mathrm{Pl}}^{2}}\ln \frac{E^{2}}{\mu
^{2}},  \label{deltal}
\end{equation}
where $m$ is the mass of the particle circulating in the loop, $E$ is the
typical energy of the process of interest and $\mu $ is a reference energy.
Taking one hundredth of an electronvolt as the mass $m_{\nu }$ of the
lightest neutrino and expressing all quantities as energies, we have 
\begin{eqnarray}
\sqrt{|\Lambda |} &\sim &10^{-42}\mathrm{GeV},\qquad \frac{m_{\nu }^{2}}{M_{%
\mathrm{Pl}}}\sim 10^{-41}\mathrm{GeV},\qquad \frac{m_{e}^{2}}{M_{\mathrm{Pl}%
}}\sim 10^{-26}\mathrm{GeV},  \notag \\
\frac{m_{\mu }^{2}}{M_{\mathrm{Pl}}} &\sim &10^{-21}\mathrm{GeV},\qquad 
\frac{m_{Z}^{2}}{M_{\mathrm{Pl}}}\sim \frac{m_{t}^{2}}{M_{\mathrm{Pl}}}\sim
10^{-15}\mathrm{GeV}.  \label{data}
\end{eqnarray}

Taking $\Lambda _{R}\sim \Delta \Lambda $ [since, by (\ref{data}), $\Lambda $
is comparable to or smaller than its radiative corrections], the corrections
due to the cosmological constant are 
\begin{equation}
\frac{|\Delta \Lambda |}{\Delta m^{2}}\ln |\Delta \Lambda |,  \label{corre}
\end{equation}
where $\Delta m$ is the energy resolution of the instruments employed in the
processes we want to study. Typical values for the experimental errors of
the particle masses are 
\begin{eqnarray}
\Delta m_{e} &\sim &10^{-13}\mathrm{GeV,\qquad }\Delta m_{\mu }\sim 10^{-9}%
\mathrm{GeV},\qquad \Delta m_{Z}\sim 10^{-3}\mathrm{GeV},  \notag \\
\Delta m_{W} &\sim &10^{-2}\mathrm{GeV},\qquad \Delta m_{t}\sim 1\mathrm{GeV}%
.  \label{reso}
\end{eqnarray}

The radiative corrections due to a particle of mass $m$ are negligible at
energies $E\ll m$, where the particle is effectively integrated out. Formula
(\ref{reso}) shows that the best energy resolution is achieved in the
measurement of the electron mass, which however does not involve high
energies. This means that, say, $m_{t}^{2}/M_{\mathrm{Pl}}$ is not relevant
to $\Delta m_{e}$.

Even if we take the ratio between the size of the universe and the Planck
distance, the logarithms of (\ref{lr}), (\ref{deltal}) and (\ref{corre})
provide a couple of orders of magnitude at most. In the end, formula (\ref%
{corre}) gives 
\begin{equation}
\frac{|\Delta \Lambda |}{\Delta m^{2}}\ln |\Delta \Lambda |\sim 10^{-20}
\label{ratio}
\end{equation}
or less, in all the relevant cases. This makes the corrections due to the
cosmological constant negligible for all practical purposes, as promised. 
\begin{figure}[t]
\begin{center}
\includegraphics[width=12truecm]{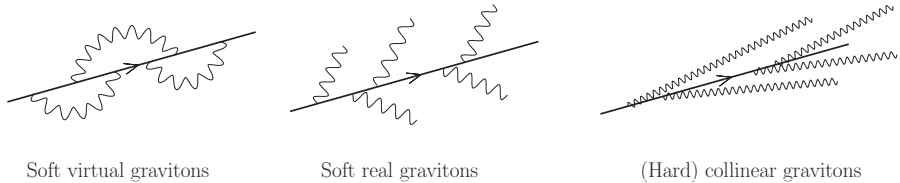}
\end{center}
\caption{Soft and collinear gravitons. Both can be virtual or real. The
collinear ones can be both soft and hard.}
\label{irdiv}
\end{figure}

To complete our analysis, we also consider the on-shell infrared divergences
(see fig. \ref{irdiv}), which are logarithmic [i.e. $\sim \ln
(m_{g}^{2}-2\Lambda )$] and of two types: soft or collinear. The soft
infrared divergences are due to gravitons of small momenta, which can be
real or virtual. The collinear divergences are due to the emissions of
gravitons at small angles with respect to the incoming or outgoing particles.

Detectors have a finite resolution $\Delta E$, which means that they cannot
distinguish a particle from a \textquotedblleft jet\textquotedblright\ made
of the particle plus gravitons of momenta smaller than $\Delta E$. From the
detector's viewpoint, all such states are equivalent, which makes it
necessary to sum over them. It can be proved \cite{infred,weinb} that the
result of the resummation over the real soft gravitons with momenta smaller
than $\Delta E$ cancels the infrared divergences due to the resummation of
the virtual soft gravitons. The net result is infrared regular and depends
on the resolution $\Delta E$. This means that the logarithms $\ln
(m_{g}^{2}-2\Lambda )$ cancel out, so there is no problem to take the limit
of vanishing gauge masses after the cancellation.

The collinear infrared divergences have similar, but also different
properties. If the angular resolution $\Delta \theta $ is finite, the
divergences due to the virtual and the real collinear gravitons mutually
cancel out and the final result depends on $\Delta \theta $. If the
collinear divergences (such as those associated with incoming particles) do
not cancel out, they can be removed by studying Altarelli-Parisi evolution
equations. So doing, they are practically buried under the initial
conditions of such equations and eliminated by means of reference
measurements. All the other measurements are then predictive and free of
collinear logarithms $\ln (m_{g}^{2}-2\Lambda )$. There, the limit of
vanishing gauge masses can be safely taken.

As said, we can keep the gauge masses nonvanishing throughout the
calculation, as long as $|m_{g}^{2}-2\Lambda |\sim |\Lambda |$.
Alternatively, we can take the limit $m_{g}\rightarrow 0$ at the end. Note
that, although the logarithm $\ln (m_{g}^{2}-2\Lambda -i\epsilon )$
generates imaginary parts for $m_{g}^{2}<2\Lambda $, the optical theorem,
expressed by the cutting equations, involves also the opposite prescription,
coming from the shadowed portions of the cut diagrams, and the final result
is real.

It is worth to stress that the diagrammatic cutting equations are satisfied
at $\Lambda <0$ as well as $\Lambda >0$, although they imply unitarity only
up to the corrections due to $\Lambda $. We might want to know whether they
imply unitarity even if we included such corrections. This issue is a bit
academic, since at the practical level it may require detectors that can
resolve wavelengths comparable to the size of the universe.

The situation is as follows. The procedure outlined so far leads to
well-defined cutting equations and tentative \textquotedblleft mathematical
cross sections\textquotedblright . If we assume that hypothetical
experiments are sensitive to the $\Lambda $ corrections, they also give
well-defined results: the numbers of events counted by their detectors. The
point is that we can no longer state that the two numbers -- the
mathematical prediction and the result of the experiment -- must coincide,
because the theory that relates them (made of asymptotic states, reduction
formulas and scattering amplitudes) is available only at $\Lambda =0$. A
problem of interpretation survives in this case.

In conclusion, we have built a theory of scattering that is unitary up to
the corrections due to $\Lambda $ and its running, which are negligible for
all practical purposes. Luckily, it is correct to expand the metric tensor
around flat spacetime, which makes the calculations doable.

\section{Quantum gravity: ultraviolet complete theory}

\label{qg2}

\setcounter{equation}{0}

In this section we generalize the proof of unitarity up to corrections due
to the cosmological constant to the ultraviolet complete theory of quantum
gravity formulated in 2017 in ref. \cite{LWgrav}, as well as its
higher-dimensional variants (listed in ref. \cite{correspondence}). We show
that the problem can be reduced to the one of the previous two sections, by
separating the Hilbert-Einstein sector from the higher-derivative sector.

A unique unitary and strictly renormalizable theory exists in four
dimensions. Its classical action, coupled to matter, can be written in two
basic ways. If we use higher derivatives, it reads 
\begin{equation}
S_{\mathrm{QG}}(g,\Phi )=-\frac{1}{2\kappa ^{2}}\int \mathrm{d}^{4}x\sqrt{-g}%
\left[ 2\Lambda +\zeta R+\alpha \left( R_{\mu \nu }R^{\mu \nu }-\frac{1}{3}%
R^{2}\right) -\frac{\xi }{6}R^{2}\right] +S_{\mathfrak{m}}(g,\Phi ),
\label{SQG}
\end{equation}%
where $\alpha $, $\xi $, $\zeta $ and $\kappa $ are real positive constants,
the reduced Planck mass is $\bar{M}_{\mathrm{Pl}}=M_{\mathrm{Pl}}/\sqrt{8\pi 
}=\sqrt{\zeta }/\kappa $, $S_{\mathfrak{m}}$ is the covariantized action of
the standard model (or an extension of it), equipped with the nonminimal
couplings required by renormalization, and $\Phi $ are the matter fields.

For various purposes, it is convenient to eliminate the higher derivatives
by adding extra fields. Then the action reads \cite{absograv} 
\begin{equation}
S_{\mathrm{QG}}^{\prime }(g,\phi ,\chi ,\Phi )=\tilde{S}_{\mathrm{HE}%
}(g)+S_{\chi }(g,\chi )+S_{\phi }(\tilde{g},\phi )+S_{\mathfrak{m}}(\tilde{g}%
\mathrm{e}^{\kappa \phi },\Phi ),  \label{SQG2}
\end{equation}
where 
\begin{eqnarray}
\tilde{S}_{\mathrm{HE}}(g) &=&-\frac{1}{2\kappa ^{2}}\int \mathrm{d}^{4}x%
\sqrt{-g}\left( 2\tilde{\Lambda}+\tilde{\zeta}R\right) ,  \notag \\
S_{\phi }(g,\phi ) &=&\frac{3\hat{\zeta}}{4}\int \mathrm{d}^{4}x\sqrt{-g}%
\left[ \nabla _{\mu }\phi \nabla ^{\mu }\phi -\frac{m_{\phi }^{2}}{\kappa
^{2}}\left( 1-\mathrm{e}^{\kappa \phi }\right) ^{2}\right] ,  \label{ss} \\
S_{\chi }(g,\chi ) &=&\tilde{S}_{\mathrm{HE}}(\tilde{g})-\tilde{S}_{\mathrm{%
HE}}(g)+\int \mathrm{d}^{4}x\left[ -2\chi _{\mu \nu }\frac{\delta \tilde{S}_{%
\mathrm{HE}}(g)}{\delta g_{\mu \nu }}+\frac{\tilde{\zeta}^{2}}{2\alpha
\kappa ^{2}}\sqrt{-g}(\chi _{\mu \nu }\chi ^{\mu \nu }-\chi ^{2})\right]
_{g\rightarrow \tilde{g}},  \notag
\end{eqnarray}
and 
\begin{eqnarray*}
\tilde{g}_{\mu \nu } &=&g_{\mu \nu }+2\chi _{\mu \nu }+\chi _{\mu \nu }\chi
-2\chi _{\mu \rho }\chi _{\nu }^{\rho },\qquad \hat{\zeta}=\zeta \left( 1+%
\frac{4}{3}\frac{\xi \Lambda }{\zeta ^{2}}\right) , \\
\tilde{\Lambda} &=&\Lambda \left( 1+\frac{2}{3}\frac{(\alpha +2\xi )\Lambda 
}{\zeta ^{2}}\right) ,\qquad \tilde{\zeta}=\zeta \frac{\tilde{\Lambda}}{%
\Lambda }.
\end{eqnarray*}

In addition to the matter fields $\Phi $, the theory describes the graviton,
through the metric tensor $g_{\mu \nu }$, a scalar field $\phi $ of squared
mass $m_{\phi }^{2}=\zeta /\xi $\ and a spin-2 field $\chi _{\mu \nu }$ of
squared mass $m_{\chi }^{2}=\tilde{\zeta}/\alpha $. Making formula (\ref{ss}%
) more explicit, it is easy to show that the $\chi _{\mu \nu }$ quadratic
action is a covariantized Pauli-Fierz action \cite{paulifierz} with the
wrong overall sign, plus nonminimal terms \cite{absograv}. This means that,
to have unitarity, the field $\chi _{\mu \nu }$ must be quantized as a
fakeon. Instead, the $\phi $ action has the correct sign, so $\phi $ can be
quantized either as a fakeon or a physical particle, leading to two
physically inequivalent theories. We recall that if the Feynman quantization
prescription is used for all the fields, the Stelle theory is obtained \cite%
{stelle}, where $\chi _{\mu \nu }$ is a ghost. In that case, unitarity is
violated at energies larger than $m_{\chi }$.

Renormalizability can be straightforwardly proved from the action (\ref{SQG}%
), because it does not depend on the quantization prescription \cite{fakeons}%
. Therefore, the beta functions coincide with those of the Stelle theory 
\cite{beta}.

Working with the action (\ref{SQG2}), equipped with the gauge-mass terms (%
\ref{deltalm}), the theory of scattering in the presence of a cosmological
constant can be formulated as explained in the previous two sections. The
massive fields $\phi $ and $\chi _{\mu \nu }$ can be viewed as additional
matter fields. As before, we can focus on the Hilbert-Einstein sector,
described by the action $\tilde{S}_{\mathrm{HE}}(g)$, and use the special
gauge (\ref{specialtwo}) to quantize the gauge-trivial poles as fakeons and
the physical poles by means of the Feynman prescription. Note that we have
two types of fakeons, here: those due to the higher-derivatives of (\ref{SQG}%
), such as $\chi _{\mu \nu }$, and those that belong to the gauge-trivial
sector, due to (\ref{specialtwo}). The former will be called hard fakeons,
while the latter will be called gauge fakeons. The $\lambda $-independent
thresholds may be physical or involve hard fakeons. If they are physical,
they are circumvented by means of the Feynman prescription. If they involve
hard fakeons they are circumvented by means of the average continuation. The 
$\lambda $ dependent thresholds always involve gauge fakeons and require the
average continuation.

The gauge mass terms (\ref{deltalm}) make the massive theory
nonrenormalizable. This does not pose obstacles to the proof of unitarity,
as we know. As far as potential infrared divergences are concerned,
nonrenormalizable terms are also not a problem, since they can only improve
the behaviors (\ref{IR}). The nonrenormalizable sector disappears altogether
when the gauge masses are sent to zero, so it is not necessary to include
new vertices at the tree level, multiplied by independent parameters, to
subtract the divergent parts that belong to that sector. Nevertheless, in
the next section we show how to modify the tree-level Lagrangian in a simple
way, to include all the counterterms we need.

\bigskip

Strictly renormalizable theories with analogous features exist in every even
dimensions $d$ greater than or equal to six \cite{correspondence}. Their
classical actions read 
\begin{equation}
S_{\mathrm{QG}}^{d}=-\frac{1}{2\kappa ^{2}}\int \mathrm{d}^{d}x\hspace{0.01in%
}\sqrt{-g}\left[ 2\Lambda +\zeta R+\hat{G}_{\mu \nu }P(D^{2})\hat{G}^{\mu
\nu }-\hat{G}P^{\prime }(D^{2})\hat{G}+\mathcal{O}(R^{3})\right] +S_{%
\mathfrak{m}}^{d}(g,\Phi ),  \label{sdqg}
\end{equation}%
where 
\begin{equation*}
\hat{G}_{\mu \nu }=\hat{R}_{\mu \nu }-\frac{1}{2}g_{\mu \nu }\hat{R},\qquad 
\hat{G}=g^{\mu \nu }\hat{G}_{\mu \nu },\qquad \hat{R}_{\mu \nu }=R_{\mu \nu
}+\frac{\Lambda }{\zeta }g_{\mu \nu },
\end{equation*}%
$S_{\mathfrak{m}}^{d}$ is the action of the matter fields $\Phi $, $P$ and $%
P^{\prime }$ denote real polynomials of degree $(d-4)/2$ and $\mathcal{O}%
(R^{3})$ collects the local Lagrangian terms that have dimensions smaller
than or equal to $d$ and are built with at least three curvature tensors and
their covariant derivatives.

As before, we separate the Hilbert-Einstein sector by introducing extra
fields. We first illustrate the procedure in a simple toy model. Consider
the Lagrangian 
\begin{equation}
L=\frac{1}{2}(\partial _{\mu }\varphi )(\partial ^{\mu }\varphi )-(\square
\varphi )Q\square \varphi -V(\varphi ),  \label{elle}
\end{equation}%
where $\varphi $ is a scalar field, $V(\varphi )$ is an interaction
potential (containing vertices that are at least cubic in $\varphi $) and $Q$
is a possibly field-dependent polynomial of the partial derivatives $%
\partial _{\mu }$. First, we add extra fields $\chi $ and $\bar{\chi}$ of
bosonic statistics and fields $\xi $ and $\bar{\xi}$ of fermionic
statistics, to rewrite the Lagrangian in the equivalent form 
\begin{equation}
L^{\prime }=\frac{1}{2}(\partial _{\mu }\varphi )(\partial ^{\mu }\varphi )+%
\bar{\chi}Q\chi -(\bar{\chi}+\chi )Q\square \varphi -\bar{\xi}Q\xi
-V(\varphi ).  \label{ellep}
\end{equation}%
The propagators of the extra fields $\chi $, $\bar{\chi}$, $\xi $ and $\bar{%
\xi}$ contain old and new poles. The old poles match poles of the $\varphi $
propagator of the initial Lagrangian (\ref{elle}) and must be quantized like
those. The new poles can be quantized with the prescription we want, as long
as it is the same for all of them, since they have to compensate one
another. For definiteness, we assume that they are quantized with the fakeon
prescription. The equivalence between $L$ and $L^{\prime }$ is easily proved
by integrating over $\chi $, $\bar{\chi}$, $\xi $ and $\bar{\xi}$, noting
that the Jacobian determinants cancel out.

The next step is to diagonalize the Lagrangian $L^{\prime }$ by means of the
field redefinition 
\begin{equation}
\varphi \rightarrow \varphi -Q(\bar{\chi}+\chi )\equiv \tilde{\varphi},
\label{fred}
\end{equation}%
which gives 
\begin{equation*}
L^{\prime }=\frac{1}{2}(\partial _{\mu }\varphi )(\partial ^{\mu }\varphi )+%
\bar{\chi}Q\chi +\frac{1}{2}(\bar{\chi}+\chi )Q\square Q(\bar{\chi}+\chi )-%
\bar{\xi}Q\xi -V(\tilde{\varphi}).
\end{equation*}%
We could introduce further fields of fermionic statistics to account for the
Jacobian determinant of the redefinition (\ref{fred}). We do not do so,
because we work with the dimensional-regularization technique, where such a
determinant is identically one. In the end, $L^{\prime }$ has a standard $%
\varphi $ quadratic action and all the higher-derivative terms act on the
extra fields.

Let us now come to the action (\ref{sdqg}). We introduce extra fields $\chi
_{\mu \nu }$, $\bar{\chi}_{\mu \nu }$, $\phi $ and $\bar{\phi}$ of bosonic
statistics, as well as extra fields $\xi _{\mu \nu }$, $\bar{\xi}_{\mu \nu }$%
, $\xi $ and $\bar{\xi}$ of fermionic statistics, to obtain 
\begin{equation*}
S_{\mathrm{QG}}^{d\hspace{0.01in}\prime }=S_{g}^{d}+S_{\chi }^{d}+S_{\phi
}^{d}+S_{\mathfrak{m}}^{d},
\end{equation*}%
where 
\begin{eqnarray*}
&&S_{g}^{d}=-\frac{1}{2\kappa ^{2}}\int \mathrm{d}^{d}x\hspace{0.01in}\sqrt{%
-g}\left[ 2\Lambda +\zeta R+\mathcal{O}(R^{3})\right] , \\
S_{\chi }^{d} &=&-\frac{1}{2\kappa ^{2}}\int \mathrm{d}^{d}x\hspace{0.01in}%
\sqrt{-g}\left[ -\bar{\chi}_{\mu \nu }P(D^{2})\chi ^{\mu \nu }+(\chi _{\mu
\nu }+\bar{\chi}_{\mu \nu })P(D^{2})\hat{G}^{\mu \nu }-\bar{\xi}_{\mu \nu
}P(D^{2})\xi ^{\mu \nu }\right] , \\
S_{\phi }^{d} &=&-\frac{1}{2\kappa ^{2}}\int \mathrm{d}^{d}x\hspace{0.01in}%
\sqrt{-g}\left[ \bar{\phi}P^{\prime }(D^{2})\phi -(\phi +\bar{\phi}%
)P^{\prime }(D^{2})\hat{G}-\bar{\xi}P^{\prime }(D^{2})\xi \right] .
\end{eqnarray*}%
Then we diagonalize the quadratic part by means of the redefinition 
\begin{equation*}
g_{\mu \nu }\rightarrow g_{\mu \nu }+\frac{1}{\zeta }P(D^{2})(\chi _{\mu \nu
}+\bar{\chi}_{\mu \nu })-\frac{1}{\zeta }g_{\mu \nu }P^{\prime }(D^{2})(\phi
+\bar{\phi})\equiv \tilde{g}_{\mu \nu }.
\end{equation*}%
Again, we do not need to take care of the Jacobian determinant of this field
redefinition if we use the dimensional regularization. The resulting action
is ready to be expanded around flat space. The $h_{\mu \nu }$ quadratic
terms are diagonal, apart from the tadpole vertex originated by the
cosmological term, which must be treated as explained in section \ref{qed}.

In the end, we manage to isolate the Hilbert-Einstein action from the rest,
so the proof of unitarity up to corrections due to the cosmological constant
works as before. The polynomials $P$ and $P^{\prime }$ must satisfy suitable
restrictions, so that the poles of the free propagators have squared masses
with nonnegative real parts. The poles with negative or complex residues, as
well as those with positive residues but complex masses, must be quantized
as fakeons. Instead, the poles with positive residues and real, positive
masses can be quantized either as fakeons or physical particles.

Super-renormalizable ultraviolet complete theories also exist (see \cite%
{LWgrav}) and can be treated similarly. They are less interesting, from the
physical point of view, because they are not unique.

\section{Massive gravitons?}

\label{massive}

\setcounter{equation}{0}

The masses we have introduced for the gauge fields and the gravitons are
artifacts to carry out the proofs of unitarity to the end. In particular, in
the case of gravity they allow us to treat the cosmological constant as
explained. At the very end, the gauge masses must tend to zero, so that
gauge invariance, Lorentz invariance and general covariance are recovered.
However, in some cases it is interesting to keep the graviton masses
different from zero and study the compatibility of such an assumption with
the experimental data. In this section, we point out that our approach does
allow us to formulate a unitary theory of massive gravitons. We also compare
it with other approaches to massive gravitons available in the literature.

Normally, it is believed that gauge invariance cannot be broken explicitly
without violating unitarity. Thanks to the fakeon prescription, used for the
quantization of the poles that belong to the gauge-trivial sector, our
construction achieves precisely that goal. In gauge theories, we have been
able to keep the gauge masses nonvanishing without renouncing unitarity,
locality and renormalizability. In gravity, so far, the graviton mass terms
we have added preserve unitary and locality, but not renormalizability. Now
we elaborate more on this issue.

We start by adding an arbitrary potential $V(\kappa h)$ (which, in some
sense, corrects the cosmological term) to the higher-derivative Lagrangian (%
\ref{SQG}). In general, $V$ is just invariant under rotations, but in
particular cases it may be Lorentz invariant. Its quadratic part is made of
the mass terms (\ref{deltalm}) or (\ref{dlminv}). The resulting action 
\begin{equation}
S_{m\mathrm{QG}}(g,\Phi )=S_{\mathrm{QG}}(g,\Phi )-\frac{1}{\kappa ^{2}}\int 
\mathrm{d}^{4}x\hspace{0.01in}V(\kappa h)  \label{sqgm}
\end{equation}%
is renormalizable by power counting (with infinitely many independent
couplings). Indeed, the renormalization of the theory is governed by a power
counting that makes $\kappa $ and $h_{\mu \nu }$ dimensionless at
high-energies, so the vertices of $V$ are multiplied by parameters of
dimension four. Since the divergent parts of the loop diagrams depend on
those parameters polynomially, and the theory contains no parameters of
negative dimensions, the counterterms generated by $V$ have the same form as
the monomials contained in $V$. Thus, if the coefficients of the $V$
monomials are independent, the action (\ref{sqgm}) is renormalizable. The
action may equally well be considered nonrenormalizable, due to the presence
of infinitely many couplings in $V$. To avoid confusion and stress that the
nonrenormalizability is of a peculiar type, we say that the action (\ref%
{sqgm}) is \ hard-renormalizable, or soft-nonrenormalizable, or almost
renormalizable.

The reason why we have not used the action (\ref{sqgm}) in the previous
section is that an analogue of the special gauge is not available at present
for the higher-derivative action $S_{\mathrm{QG}}$, which means that (\ref%
{sqgm}) leads to extremely involved propagators.

Starting from (\ref{sqgm}), it is still convenient to switch to the
non-higher-derivative form of the action by inserting the extra fields $\phi 
$ and $\chi _{\mu \nu }$ explicitly. Mimicking the steps of ref. \cite%
{absograv}, the metric redefinition is 
\begin{equation*}
g_{\mu \nu }\rightarrow (g_{\mu \nu }+2\chi _{\mu \nu }+\chi _{\mu \nu }\chi
-2\chi _{\mu \rho }\chi _{\nu }^{\rho })\mathrm{e}^{\kappa \phi }.
\end{equation*}%
After the redefinition, the action we get is $S_{\mathrm{QG}}^{\prime }$
plus the transformed potential, which contains, among the other things,
extra mass terms for $\phi $ and $\chi _{\mu \nu }$ and off-diagonal
quadratic terms.

The quadratic corrections can be included into modified propagators, by
means of a resummation like (\ref{masspr}). First, it is convenient to resum
the $h$-$h$ mass terms, following the guidelines explained below formula (%
\ref{masspr}). Second, it is convenient to resum the corrections to the $%
\chi $-$\chi $ mass terms. Since such corrections are generically not of the
Pauli-Fierz type, they turn on a new pole in the $\chi $ propagator, which
describes a scalar field $\pi $. The residue of the $\pi $ pole is positive
(because the $\chi $ Pauli-Fierz action is multiplied by the wrong sign), so 
$\pi $ can be quantized as a physical particle or a hard fakeon. Third, we
resum the $\phi $-$\phi $ terms, which just correct the $\phi $ mass.
Finally, we resum the off-diagonal $h$-$\phi $-$\chi $ mass terms following
the lines explained below formula (\ref{masspr}). The physical poles, as
well as those associated with the hard fakeons, remain $\lambda $
independent up to corrections due to the gauge masses. The poles associated
with the gauge fakeons remain $\lambda $ dependent. As usual, the physical
thresholds are treated by means of the Feynman prescription, while the fake
thresholds are treated by means of the average continuation. We recall that
coinciding thresholds must be treated as limits of distinct thresholds.

Ultimately, we obtain a local, unitary and almost renormalizable theory of
massive gravitons. The theory violates general covariance and, in
particular, Lorentz symmetry. We can reduce the effects of the Lorentz
violation by choosing a Lorentz invariant potential $V$ and sending $\lambda 
$ to one after the computations of the loop diagrams. Then, the surviving
Lorentz violations start from one loop.

Due to the violation of general covariance, the gauge fakeons have physical
effects, like the hard fakeons. It is known that causality is violated at
energies larger than the fakeon masses \cite{UVQG,absograv,causalityQG,flrw}%
. To make the violations small and compatible with the data, the masses of
the hard fakeons must be large, while the potential $V$ must be small. The
latter condition follows from the fact that the gauge fakeons do not
contribute to the physical quantities in the limit of vanishing $V$ (since
they get buried into the gauge-trivial sector). A simple way to see this is
to note that the physical quantities are gauge independent in that limit,
while the sector of the gauge fakeons is $\lambda $ dependent. It is easy to
check [see formulas (\ref{lpm}) and (\ref{pp}) below] that the $\pi $ mass
is large when the gauge masses are small, so if $\pi $ is quantized as a
fakeon, the violation of causality due to it is small, as desired.

\bigskip

One of the reasons why the theories of massive gravitons attract interest is
that in many cases they remove the van Dam-Veltman-Zacharov (vDVZ)
discontinuity \cite{VDVZ}. The massive theories we have just built also
achieve this goal, since the limit of vanishing gauge masses is smooth. Not
only: the fakeon quantization prescription provides a further, Lorentz
invariant option. Consider the Pauli-Fierz Lagrangian 
\begin{equation}
L_{\mathrm{PF}}=\frac{1}{2}\left[ \partial _{\rho }h_{\mu \nu }\partial
^{\rho }h^{\mu \nu }-\partial _{\rho }h\partial ^{\rho }h+2\partial _{\mu
}h^{\mu \nu }\partial _{\nu }h-2\partial _{\mu }h^{\rho \nu }\partial _{\rho
}h_{\nu }^{\mu }-m^{2}(h_{\mu \nu }h^{\mu \nu }-h^{2})\right]   \label{LPF}
\end{equation}%
for a symmetric tensor $h_{\mu \nu }$ and add the unconventional correction 
\begin{equation}
L_{m}^{\prime }=-\frac{3m^{4}}{4(2m^{2}+\bar{m}^{2})}h^{2}  \label{lpm}
\end{equation}%
to its mass term. The propagator $-i\langle h_{\mu \nu }(k)\hspace{0.01in}%
h_{\rho \sigma }(-k)\rangle $ of the resulting Lagrangian 
\begin{equation}
L_{\mathrm{PF}}+L_{m}^{\prime }  \label{lpfp}
\end{equation}%
is the sum of the Pauli-Fierz one, which describes a spin-2 particle of mass 
$m$, plus 
\begin{equation}
-\frac{1}{6(k^{2}-\bar{m}^{2})}\left( \eta _{\mu \nu }+2\frac{k_{\mu }k_{\nu
}}{m^{2}}\right) \left( \eta _{\rho \sigma }+2\frac{k_{\rho }k_{\sigma }}{%
m^{2}}\right) ,  \label{pp}
\end{equation}%
which has a scalar pole of mass $\bar{m}$. Since the residue of this pole in
negative, the Feynman quantization prescription turns it into the
Boulware-Deser (BD) ghost \cite{boulware} and violates unitarity. However,
if we quantize the pole (\ref{pp}) as a fakeon, unitarity holds and the
vDVZ\ discontinuity is removed in a Lorentz invariant way. At the
observative level $\bar{m}$ should be small and since $1/\bar{m}$ is the
range of the violation of causality, the theory with (\ref{pp}) as a fakeon
has the problem of explaining why that violation is not observed.

Rubakov has shown that (if we do not make use of fakeons) the vDVZ
discontinuity can be removed in a unitary way if Lorentz invariance is
broken and certain restrictions on the masses are imposed \cite{rubakov}. If
Rubakov's would-be ghosts are quantized as fakeons, it might be possible to
relax some of the Rubakov conditions on the masses.

Like Rubakov's theory, the theories of massive gravitons of the previous two
sections are not Lorentz invariant. The main difference between our
propagators and the Rubakov's ones are that we have included the
gauge-fixing terms, to ensure a smooth limit of vanishing $V$, where Lorentz
invariance and general covariance are recovered.

The Dvali-Gabadadze-Porrati (DGP) model \cite{dgp} overcomes the vDVZ\
discontinuity by obtaining four-dimensional gravity from a five-dimensional
theory. In the de Rham-Gabadadze-Tolley (dRGT) model \cite{drgt} the
Boulware-Deser ghost is removed by adding nonderivative interaction terms
for the metric $g_{\mu \nu }$, which requires to introduce an extra,
reference metric $f_{\mu \nu }$. The extra metric is also present in our
model: it is the flat-space metric used to build the potential $V$.

Finally, we stress that the theory of massive gravitons (\ref{sqgm}) is
unitary (up to corrections due to the cosmological constant) and almost
renormalizable. The problem of renormalizability remains open in the
theories of refs. \cite{rubakov,dgp,drgt}, as well as in any Lorentz
invariant massive theory with quadratic Lagrangian (\ref{lpfp}).

\section{Conclusions}

\label{conclusions}

\setcounter{equation}{0}

In this paper we have worked out simple proofs of perturbative unitarity in
gauge theories and quantum gravity. The special gauge allows us to separate
the physical poles, which are quantized by means of the standard Feynman
prescription, from the poles that belong to the gauge sector, which are
quantized by means of the fakeon prescription. Inside the loop diagrams, the
dependence on a gauge-fixing parameter $\lambda $ allows us to distinguish
the physical thresholds, which are overcome analytically, from the fake
thresholds, which are overcome non analytically by means of the average
continuation. The proof works for nonrenormalizable and ultraviolet complete
theories.

We also clarified a number of nontrivial issues about the formulation of the
theory of scattering in the presence of a cosmological constant. The
scattering amplitudes are defined by expanding the metric around flat space.
They obey unitarity up to corrections due to the cosmological constant,
which can be neglected in all practical situations. In the (unrealistic)
case that such corrections became important, the cutting equations hold, but
their physical meaning remains unclear.

We have introduced nonvanishing gauge masses for various practical purposes.
Gauge invariance, Lorentz invariance and general covariance are recovered in
the limit where the gauge masses vanish. If we keep the gauge masses
different from zero, our construction provides a way of building local,
unitary and almost renormalizable theories of massive gauge fields and
gravitons (which violate Lorentz invariance, gauge invariance and general
covariance). Usually, it is believed that the explicit breaking of a gauge
symmetry leads to the violation of unitarity. We have shown that once
fakeons are employed, unitarity and gauge invariance are ultimately
disentangled from each other and it is possible to break the latter without
breaking the former.

The theories with fakeons violate causality at energies larger than the
fakeon masses. The masses of the hard fakeons must be large enough, to have
compatibility with the data, while the gauge masses must be small enough,
because in the limit where they vanish the gauge fakeons do not contribute
to the physical quantities.

\vskip12truept \noindent {\large \textbf{Acknowledgments}}

\vskip 2truept

I am grateful to U. Aglietti, M. Bochicchio, D. Comelli and F. Nesti for
helpful discussions.

\end{document}